\documentclass[prb,twocolumn,amsmath,amssymb,floatfix]{revtex4}
\usepackage{graphicx}

\newcommand{\bfA}{{\bf A}}
\newcommand{\bfB}{{\bf B}}
\newcommand{\be}{\begin{eqnarray}}
\newcommand{\ee}{\end{eqnarray}}
\newcommand{\da}{\dagger}
\newcommand{\bfr}{{\bf r}}

\newcommand{\oncite}{\onlinecite}

\begin{document}

\title{Quantum phase diagrams of fermionic dipolar gases for an arbitrary orientation of dipole moment in a planar array of 1D tubes}

\author{Yi-Ping Huang$^1$ and Daw-Wei Wang$^{1,2}$} 
\affiliation{$^1$ Physics Department, National Tsing-Hua University, Hsinchu, Taiwan 300, ROC
\\
$^2$ Physics Division, National Center for Theoretical Sciences,
Hsinchu, Taiwan 300, ROC}

\date{\today}

\begin{abstract}
We systematically study ground state properties of fermionic dipolar gases
in a planar array of one-dimensional potential tubes for an
arbitrary orientation of dipole moments. Using the Luttinger liquid theory with the generalized Bogoliubov transformation, we calculate the elementary excitations and the Luttinger scaling exponents for various relevant quantum orders. The complete quantum phase diagrams for arbitrary polar angle of the dipole moment is obtained, including charge density wave, $p$-wave superfluid, inter-tube gauge-phase density wave, and inter-tube $s$-wave superfluid, where the last two breaks the $U(1)$ gauge symmetry of the system (conservation of particle number in each tube) and occurs only when the inter-tube interaction is larger than the intra-tube interaction. We then discuss the physical properties of these many-body phases and their relationship with some solid state systems.
\end{abstract}

\maketitle
\section{Introduction}

It is well-known that the system of ultracold atoms has been the most experimentally flexible system to study the strongly correlated phenomenons in the recent decade, due to the widely tunable interaction strength, dimensionality and/or the species of underlying atoms. Since the many-body physics in one-dimensional systems has been extensively studied by various analytic and numerical methods in the last few decades, it is therefore important to compare the theoretical predictions to the experimental observation in the recent low-dimensional cold atom systems. Several important works along this direction have been reported recently, including 
the Tonks-Girardeau gas [\oncite{Bloch and the others}], Luttinger liquid (LL) behavior [\oncite{cazallila}], two-component fermionic gas [\oncite{two}], 1D BEC-BCS crossover [\oncite{crossover}], polaronic effects in 
Bose-Fermi mixture [\oncite{mathey_wang}], and 1D spinor gases [\oncite{spinor}] etc.. However, due to the short-range nature of
atomic interaction, it is usually not easy to study how the interaction 
between particles in different 1D tubes can bring different many-body effects, which, however, can be easily achieved in the traditional solid state system due to the long-ranged nature of Coulomb interaction.
Knowing the fact that the dipolar interaction between spinful atoms and polar molecules can be also prepared in various circumstances [\oncite{Cr,JILA_fermions}], it is therefore getting more and more attention on the effect of long-ranged dipolar interaction between ultracold atoms/molecules in low-dimensional systems [\oncite{wang_dipolar_liquid_Santos,Benjamin2009}]. 
Among these earlier works, we are especially interested in the systems of multi-tubes [\oncite{Chang2008,Quintanilla2009,cazallila}],
which provides a direct example to study how the quasi-1D system can be merge to the physics of two-dimensional system.

\begin{figure}
\includegraphics[width=8cm]{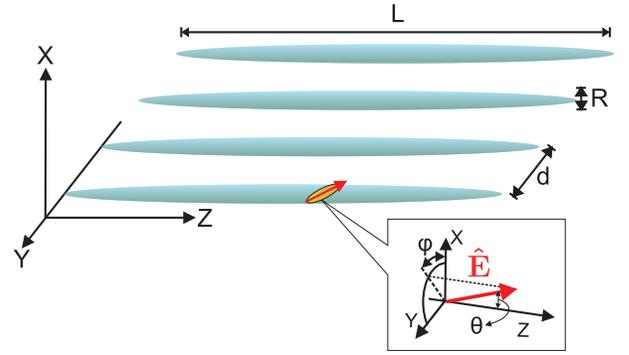}
\caption{(Color online) A planar array of 1D tubes of a dipolar gas with the inter-tube distance, $d$, the tube radius, $R$, and the tube length of tube, $L$. In this paper, we assume $L\gg d\gg R$, and therefore the system can be regarded as a quasi-1D system. The dipole orientation is along the direction of the external field, $\hat{E}$, which is tilted with a polar angle $\theta$ from the $z$ axis. The azimuthal angle is denoted by $\phi$.}
\label{system}
\end{figure}
In this paper we extend our earlier work in the study of double-tube systems [\oncite{Chang2008}] to investigate the ground state properties of a planar an array of 1D tubes loaded with dipolar fermionic atoms/molecular, assuming the inter-tube tunneling is energetically negligible. The dipole moments are polarized by an external electric or magnetic field in an arbitrary direction (see Fig. \ref{system}), providing a variety of interaction matrix elements that are not available in traditional solid state systems.
Using the multi-component Luttinger liquid theory, we then diagonalize the effective system Hamiltonian exactly, and obtain the quasi-long ranged quantum orders by studying the correlation functions. We further study the elementary excitations and investigate the regime of system instability via a mode softening in the long wave-length limit. 

Our results show that, some unexpected inter-tube correlation or pairing mechanism can be found when the dipole moment is tilted near a magic angle, $\theta_c=\cos^{-1}\sqrt{1/3}$ (see Fig. \ref{system}), where the intra-tube interaction is relatively smaller than the inter-tube interaction. Depending the inter-tube interaction is positive or negative (for different angle, $\phi$, see Fig. \ref{system}), fermions in the neighboring tubes can have inter-tube gauge-phase density wave (GPDW) or inter-tube $s$-wave superfluid ($s$-SF), breaking the $U(1)$ gauge symmetry (i.e. particle number conservation in each tube) of the system (in the quasi-long-ranged order sense). We note that in the quantum Hall double layer system, the gauge symmetry breaking phase (or called inter-layer coherence, pseudo-spin ferromagnetism, or exciton condensate) has been observed experimentally in a semi-conductor heterostructure [\oncite{quantum_Hall}], showing an interaction-enhanced single particle zero-bias tunneling current. We believe similar results may be also observed in the multi-tube system we discuss here. For the inter-tube $s$-wave superfluid state, we expect a perfect transmission for a particle moving through a potential barrier along the longitudinal direction of the tubes at zero temperature, and a power-law decaying behavior at finite temperature, similar to the results predicted by Kane and Fisher [\oncite{Kane}] on a single interacting 1D electronic gas. When the angle $\theta$ is larger or smaller than $\theta_c$, intra-tube interaction becomes dominant and the system becomes simple charge density wave (CDW) or intra-tube $p$-wave superfluid ($p$-SF) phases, consistent with the standard mean field results. (We will show a clear definition and physical picture for each of these phases later.) Our works therefore provide a clear evidence of the strong dipolar long-ranged interaction effects in this quasi-1D systems. 

This paper is organized as following:
In Sec. \ref{s1}, we present the interaction matrix element, the system Hamiltonian in Luttinger liquid theory, and the diagonalization method for 
the multi-component Luttinger liquid Hamiltonian. We then study the elementary excitations in Sec. \ref{s2} and the correlation functions in Sec. \ref{s3}, where some details of calculation are shown in Appendix \ref{B}. We then show the complete quantum phase diagram for arbitrary dipole angle and discuss related physical properties in Sec. \ref{s4}. We then summarize our results in Sec. \ref{s5}.

\section{System Hamiltonian and diagonalization method}
\label{s1}

\subsection{Interaction matrix element}

The multi-tube system considered in this paper is shown
in Fig. \ref{system}, where the trapping potential can be
generated by a 2D optical lattice in the $x-y$ plane with a stronger lattice potential in the $x$-axis. Additional
weak magneto-optical potential is applied in the longitudinal direction
($z$) to trap dipolar atoms/molecules in such quasi-1D potential. In this paper, we assume the longitudinal length scale, $L$, is much larger than the
inter-tube distance $d$ and tube radius $R$ (i.e. $L\gg d\gg R$).
For simplicity, we assume the inhomogeneity of the weak trapping
potential along the $y$ and $z$ axises can be neglected, and therefore the system can be regarded as an array of identical 1D tubes with total number of tubes, $N$. 

Throughout this paper, we always consider the case when dipolar atoms/molecules are all loaded in the lowest subband of each tube, which has a transverse confinement wavefunction:
$\phi_j(x,y)=\frac{1}{\sqrt{\pi}R}e^{-(y^2+(y-jd/2)^2)/2R^2}$.
The resulting effective 1D system Hamiltonian then
can be written to be $H=H_0+H_I$, where $H_0$ is the kinetic energy:
\be
H_0=\sum_{j=1}^N\int_0^L dz\psi^\dagger_j(z)
\frac{-\hbar^2}{2m}\partial_z^2\psi_j^{}(z)
\label{H_0}
\ee
with $m$ being the mass of dipolar particles and $\psi_j(z)$ being the fermion field operator in the $j$th tube, and $H_I$
is the interaction energy: 
\be
H_I&=&\frac{1}{2}
\sum_{j,j'}\int_0^L dz\int_0^L dz'V_{|j-j'|}(z-z')
\nonumber\\
&&\times\psi^\dagger_j(z)\psi^{}_j(z)
\psi^\dagger_{j'}(z')\psi^{}_{j'}(z').
\label{H_I}
\ee
Here $V_{|j-j'|}(z)$ is the dipolar interaction 
between molecules in the same tubes ($j=j'$) or in different 
tubes ($j\neq j'$), obtained by integrating out the transverse
degree of freedom ($\bfr_\perp\equiv (x,y)$):
\be
V_{|j-j'|}\left(z\right)&=&
\int d{\bf r}_{1,\perp}\int d{\bf r}_{2,\perp}
\left|\phi_j(\bfr_{1,\perp})\right|^2
\left|\phi_{j'}(\bfr_{2,\perp})\right|^2
\nonumber\\
&&\times V_{d}\left({\bf r}_1-\bfr_2\right),
\label{V_bare}
\ee
where $V_{d}(\bfr)=D^2(1-3(\hat{r}\cdot\hat{E})^2)/|\bfr|^3$ 
is the bare dipolar interaction with $D$ being the electric dipole moment in the c.g.s unit. $\hat{E}$ is the unit vector along the external field, parallel to the direction of dipole moment (see Fig. \ref{system}).
Since in general the electric dipole interaction is much stronger (and tunable) than the magnetic dipole interaction, in the the rest of
this paper we will use polar molecules as the underlying particles 
for further discussion. Extension of our results to the magnetic dipolar atoms is straightforward.

\begin{figure}
 \includegraphics[width=8cm]{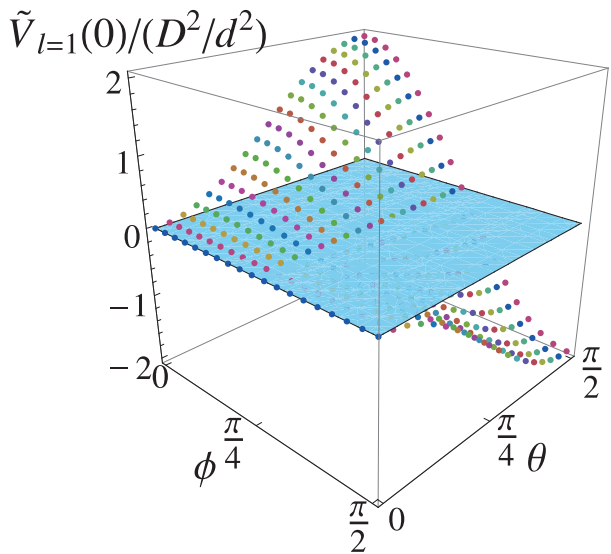}
\caption{(Color online) The inter-tube interaction for two neighboring tubes, $|j-j'|=1$, with arbitrary dipole orientation and $d/R=5$, see Eq. (\ref{k0inter_a}).}
\label{inter_a}
\end{figure}

For the convenience of later study, here we define the interaction matrix element to be the Fourier transform of the bare dipole interaction in momentum space, i.e. we take the first Born approximation for the two-particle scattering for simplicity (see Ref. [\oncite{3d_dipole_wang}]):
we have $\widetilde{V}_{|j-j'|}(q)\equiv\int dz V_{|j-j'|}(z)\,e^{-iqz}$. Although one has to evaluate $\widetilde{V}_{|j-j'|}(q)$ numerically for finite value of $q$, we can still obtain an analytic form of their zero momentum ($q=0$) values, which dominates the low energy physics in the Luttinger liquid theory. By integrating over the transverse confinement wavefunction, we have
\be
\widetilde{V}_{0}\left(0\right)=\frac{D^2(1-3\cos^2\theta)}{R^2}
\label{k0intra}
\ee
for the intra-tube interaction and 
\be
\widetilde{V}_{l}\left(0\right)&=&
\frac{D^2\cos2\phi\sin^2\theta}{l^2d^2}
\nonumber\\
&&\times\left\{2-\left(2+\left(\frac{ld}{R}\right)^2\right)
\exp\left[-\frac{l^2d^2}{2R^2}\right]\right\}
\label{k0inter_a}
\ee
for the inter-tube interaction ($l\equiv |j-j'|$).
It is very easy to see that when $\theta=\theta_c=\cos^{-1}\sqrt{1/3}$
the intra-tube interaction becomes zero and change sign, while the inter-tube interaction vanishes and changes sign at $\phi=\pi/4$. In other words, we can have a wide parameter range to tune the sign of interaction matrix element by tuning the direction of external field to explore various kinds of interesting many-body physics. In Fig. \ref{inter_a}, we show the value of nearest neighboring inter-tube interaction matrix element, $\widetilde{V}_1(0)$, as a function of $\theta$ and $\phi$. We note that such interesting kind of inter-tube interaction matrix element cannot be realized in the traditional solid state system and therefore may bring some physics even not predicted before in the condensed matter theory.

\subsection{Luttinger liquid Hamiltonian}
\label{s1.1}

\begin{figure}
 \includegraphics[width=8cm]{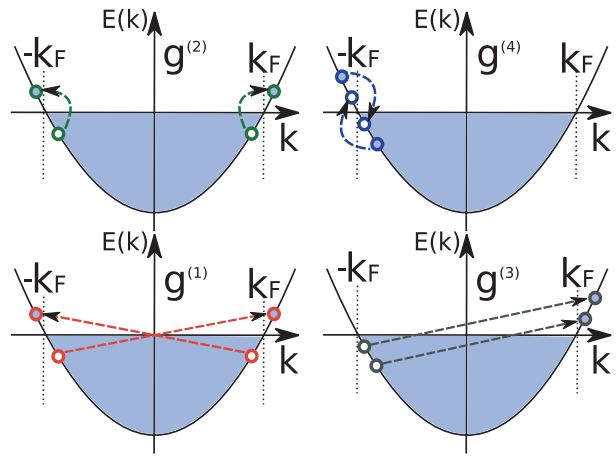}
\caption{(Color online) Different scattering process for Tomonaga-Luttinger model. We assume the system is away from half filling regime and consider 
the small momentum transfer only, so that $g^{(1)}$ and $g^{(3)}$ scatterings are neglected in our system. Therefore, we only consider 
forward scattering, $g^{(2)}$ and $g^{(4)}$, in our system Hamiltonian.}
\label{LL}
\end{figure}

For one-dimensional non-interacting fermions, the elementary 
particle-hole excitations around the two Fermi points can be 
well-approximated by the linearized dispersion, known as the
Tomonaga-Luttinger model [\oncite{Tomonaga1950,Luttinger1963,Haldane}].  
In the low energy limit of an interacting system (i.e. when the characteristic energy scale is much
smaller than the Fermi energy), the most important excitations
are still around the Fermi points and therefore the ground state and 
the elementary excitation properties can be well-described by 
the so-called Luttinger liquid theory [\oncite{Haldane,Solyom1979,Voit}],
where the single particle excitation of the Landau-Fermi liquid becomes totally destroyed by the inter-particle interaction, leaving only gapless collective modes as the low energy elementary excitations.
Within the standard Luttinger liquid theory, only four kinds of 
scattering near the two Fermi points are important (see Fig. \ref{LL}): 
$g^{(2)}$ and $g^{(4)}$ are the forward scattering without changing the 
direction of scattered fermions, $g^{(1)}$ is the backward scattering with the change of momentum, $2k_F$, and $g^{(3)}$ is the Umkalpp scattering, which is important only when a lattice potential is present and the filling fraction is close to unit. In our present work, we are interesting in the regime when $g^{(1)}$ is negligible or irrelevant, and $g^{(3)}$ can be omitted since no lattice potential along the longitudinal direction of 1D tubes is considered here. Renormalization group for these multi-tube systems
can be used to study the relevance of the backward scattering term (i.e. $g^{(1)}$), and the results when it becomes relevant will be presented in another place in the future [\oncite{future}]. Throughout this paper, we are interested in the regime when this backward scattering is irrelevant or at least negligible.

In the multi-tube system we consider here, the two scattered particles
can be either in the same tube (intra-tube interaction) or in two
different tubes (inter-tube interaction). 
As a result, labeling the single particle quantum states by the tube
index, $j$ ($j=1,\cdots, N$), and the left/right mover index near the 
two Fermi points ($r\equiv\pm$), the system Hamiltonian of Eqs. (\ref{H_0})
and (\ref{H_I}) can be written as $2N$ component Luttinger liquid model:
\be
H_{LL}&=&\sum_{j}H_{j}^{(0)}+\sum_{j,j'}\left(H_{j,j'}^{(2)}+H_{j,j'}^{(4)}
\right)
\ee
where $j,j'$ are the tube index. The noninteracting Hamiltonian, $H_j^{(0)}$, can be approximated by a linearized band structure about the Fermi points:
\be
H_{j}^{(0)}&=&\sum_{r=\pm}\sum_q v_F(rq-k_F)
c^{\dagger}_{j,r,q}c_{j,r,q}
\nonumber\\
&=&\frac{2\pi}{L}\sum_{q>0,r=\pm}
v_F\rho_{j,r}(q)\rho_{j,r}(-q).
\label{H_0_2}
\ee
Here $c_{j,r,q}$ is the Fermionic creation operator for dipoles with momentum $q$. $v_F$ and $k_F$ are the Fermi velocity and Fermi momentum, and are assumed to be the same for all tubes. In the last line, we have used the standard result of Luttinger model, where the kinetic energy of a linearized dispersion can be composed of two density operators [\oncite{Mattis1965}], $\rho_{j,r}(q)$, where
\be
\rho_{j,r}(q)\equiv\sum_k c_{j,r,k+q}^{\dagger}c_{j,r,k}.
\ee
Note this density operator here obeys a boson-like commutation relation,
\be
\left[\rho_{\lambda}(-q),\rho_{\lambda'}(q')\right]&=&\delta_{q,q'}
\delta_{\lambda,\lambda'}\frac{rqL}{2\pi}
\nonumber\\
&=&\delta_{q,q'}\delta_{\lambda,\lambda'}\frac{qL}{2\pi}{\rm sign}(v_\lambda),
\label{commun}
\ee
where we define $\lambda=(j,r)$ to label the tube and charilty indecies, and $v_\lambda\equiv rv_F$. As a result, $v_\lambda=\pm v_F$ for the right/left-moving channel. Such convention can be very useful for our latter application to the generalized Bogoliubov transformation. 

As for the interaction Hamiltonian of forward scattering, it is easy to show that it can be also rewritten to be the following form within the Luttinger liquid theory:
\be
H_{j,j'}^{(2)}&=&\frac{1}{L}\sum_{q}g_{j,j'}^{(2)}(q)\rho_{j,+}(q)\rho_{j',-}(-q)
\label{H_I_2}
\\
H_{j,j'}^{(4)}&=&\frac{1}{2L}\sum_{q,r=\pm}g_{j,j'}^{(4)}(q)\rho_{j,r}(q)\rho_{j',r}(-q).
\label{H_I_4}
\ee
Note that in the limit of weak interaction (i.e. interaction energy is smaller than the Fermi energy) the two kinds of forward scattering ($g^{(2)}$ and $g^{(4)}$) are the same as the bare interaction, i.e.
\be
g^{(2)}_{j,j'}(q)&=&g^{(4)}_{j,j'}(q)=\widetilde{V}_{|j-j'|}(q).
\ee
Since both the noninteracting Hamiltonian and the interaction Hamiltonian are now quadratic in terms of density fluctuation operator, the total Luttinger liquid Hamiltonian, Eqs. (\ref{H_0_2}), (\ref{H_I_2}) and (\ref{H_I_4}), can be diagonalized by using a generalized Bogoliubov transformation as described below.

\subsection{The Generalized Bogoliubov transformation}
\label{s1.2}

To make the notation more compact for the Bogoliubov transformation
of a general $N$-tube system, we use $\lambda$ to denote both
the tube index ($j$) and the chirality index ($r=\pm$).
The whole Hamiltonian shown above can be described to be
\be
H=\frac{2\pi}{L}\sum_{\lambda,\lambda',q>0}A_{\lambda,\lambda'}(q)\rho_{\lambda}
(q)\rho_{\lambda'}(-q),
\label{Hamiltonian}
\ee
where
\be
A_{\lambda,\lambda'}(q)=|v_\lambda|\delta_{\lambda,\lambda'}
+\frac{\tilde{V}_{\lambda,\lambda'}(q)}{2\pi}
\ee
is the matrix element of the matrix $\bfA$.

We diagonalized Eq. (\ref{Hamiltonian}) via the generalized Bogoliubov 
transformation developed by Penc and S\`olyom [\oncite{Penc1993}]. 
Defining the eigenstate density fluctuation to be 
$\tilde{\rho}_n(q)$ with the index $n=1,2,\cdots, 2N$, the transformation can be written to be 
\be
\tilde{\rho_n}(q)=\sum_{\lambda}w_{n,\lambda}\rho_{\lambda}(q),
\label{transform1}
\ee
where $w_{n,\lambda}$ is the matrix element of a canonical transformation
(i.e. the generalized Bogoliubov transformation) matrix. 
After the transformation, the Hamiltonian in Eq. (\ref{Hamiltonian}) should be expressed to be
\be
H&=&\frac{2\pi}{L}\sum_{q>0}\sum_n|u_n|\tilde{\rho_n}(q)\tilde{\rho_n}(-q)
\nonumber\\
&=& \frac{2\pi}{L} \sum_{q>0,n}\sum_{\lambda,\lambda'}
w_{n,\lambda}\rho_{\lambda}(q) |u_n| w_{n,\lambda'}\rho_{\lambda'}(-q) 
\label{B_diag}
\label{H2}
\ee
In other words, comparing Eq. (\ref{Hamiltonian}) and Eq. (\ref{H2}), we must have
\be
A_{\lambda,\lambda'}=\sum_n|u_n|w_{n,\lambda}w_{n,\lambda'}
\ee
or
\be
{\bf A}=\sum_{n}|u_n||w^{(n)}\rangle\langle w^{(n)}|,
\ee
where $|w^{(n)}\rangle$ is a vector with elements, $w_{n,\lambda}$.
The commutation relation of the new density operators can be also expressed to be (following Eq. (\ref{commun})):
\be
[\tilde{\rho}_n(-q),\tilde{\rho}_{n'}(q')]=\delta_{q,q'}\frac{qL}{2\pi}\langle w^{(n)}|\bfB|w^{(n')}\rangle,
\ee
where the matrix ${\bfB}$ in these two different basis should be:
\be
B_{\lambda,\lambda'}&=&\delta_{\lambda,\lambda'}{\rm sign}(v_{\lambda})
\\
B_{n,n'}\equiv\langle w^{(n)}|\tilde{B}|w^{(n')}\rangle
&=&\delta_{n,n'}{\rm sign}(u_{n}).
\label{B_orth}
\ee
From the above derivation, we want to find a basis $\{|w^{(n)}\rangle\}$ to diagonalize the matrix $\tilde{A}$ with positive eigenvalues, $|u_n|$, and the eigenvectors have to satisfy the special orthogonal relation in Eq. (\ref{B_orth}). It is easy to see that it is equivalent to solve the standard eigenvalue problem below:
\be
\bfA\bfB|w^{(n)}\rangle=u_n|w^{(n)}\rangle
\ee
with the condition of Eq. (\ref{B_orth}). This is the so-called generalized Bogoliubov transformation for the multi-component Luttinger liquid system, first developed by Penc and S\`olyom [\oncite{Penc1993}].

\section{Elementary excitations and system instability}
\label{s2}

\subsection{Results for $\phi=0$}
\label{s2.1}

The first physical properties we want to study is the elementary
excitations of such multi-tube system.
For the convenience of later discussion, here we introduce several
dimensionless parameters to describe the systems: 
$\gamma\equiv\frac{m D^2}{\hbar^2 d}$ is to measure the dipolar
interaction strength,
$d_R\equiv d/R$ is the inter-tube distance compared to the tube radius, 
and $k_F d$ is to measure the particle density in each 
tube. Since the long wave-length intra-tube interaction can be tuned 
to zero as $\theta\sim\theta_c=\cos^{-1}\sqrt{1/3}$ (see Eq. (\ref{k0intra})), it is reasonable to investigate the results for $\theta=\theta_c$ first. We also set $\phi=0$, i.e. the dipole moment is in the $x-z$ plane first for the convenience of study. Results for a more general angular parameter regime will be presented later. 
\begin{figure}
\includegraphics[width=8cm]{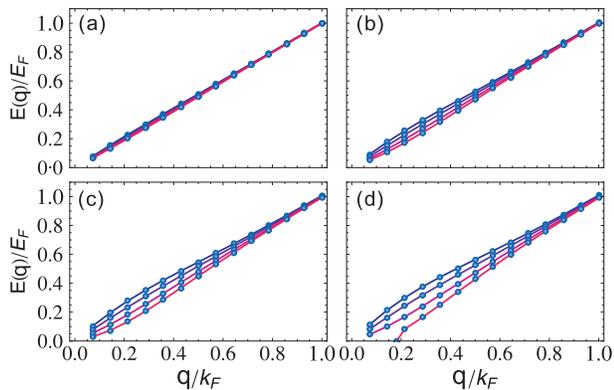}
\caption{The calculated energy dispersion for the eigen-mode excitations of
four tubes system. (a)-(d) are for $\gamma=2,6,10$, and 14 respectively,
and we set $k_Fd=7$, $d_R=5$, $\theta=\theta_c$, and $\phi=0$ here. 
The instability occurs at the lowest branch for $\gamma=14$. }
\label{dispersion}
\end{figure}
In Fig. \ref{dispersion}, we show the calculated dispersion of collective 
modes for different values of interaction strength, $\gamma$, in the
system of four tubes ($N=4$). It is easy to see that since the intra-tube
interaction is zero at $\theta=\theta_c$, the dispersion of the 
collective excitation becomes hybridized by the inter-tube interaction,
which lifts the degeneracy and changes the excitation dispersion dramatically in the long wavelength limit, while it becomes less important in the short wavelength (large momentum) regime. The upper two branches exhibit convex shape at small momentum and the lower two branches show a concave shape. This train is enhanced in strong dipole-dipole interaction regime, and the lowest energy branch eventually becomes softened
when the dipole strength, $\gamma$ is above a critical value.
By investigating the eigenstate wavefunction of the lowest energy collective
mode, we find that the instability should be related to some kinds
of phase separation state, which is derived by the strong inter-tube repulsive interaction in such multi-component system.

\begin{figure}
\includegraphics[width=8cm]{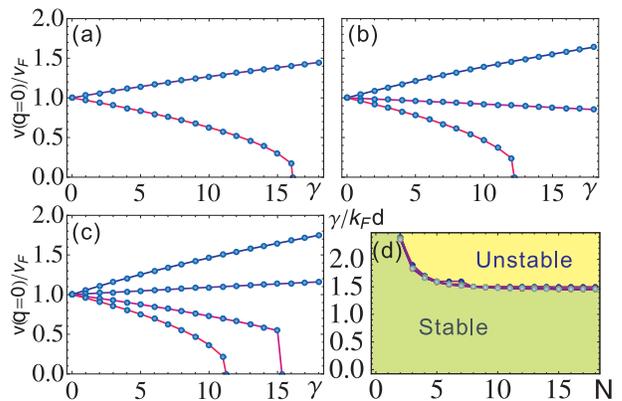}
\caption{(Color online) (a)-(c) show the group velocity at $q=0$ as a function 
of interacting strength $\gamma$ for $N=2$, 3, and 4 respectively. Here we set $d_R=5$, $\theta=\theta_c$, and $k_F d=7$. 
(d) shows the critical value, $\gamma_c$, of the softening of the lowest energy excitation as a function of numbers of tubes, $N$. Results for $k_Fd=1,2,\cdots 7$ are shown together and found scaled to a single universal curve, above which the system becomes unstable.}
\label{stability}
\end{figure}
\begin{figure}
\includegraphics[width=8cm]{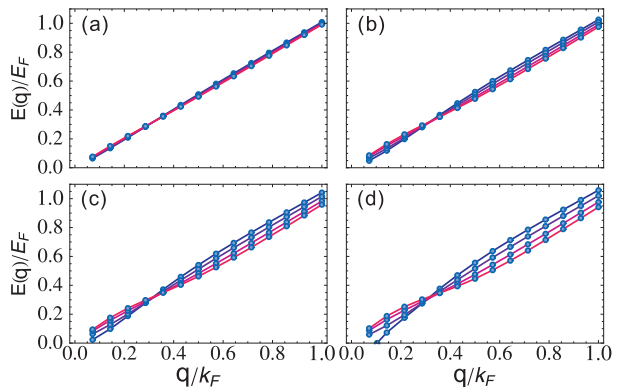}
\caption{Elementary excitation energy dispersion for $\theta=\theta_c$, $\phi=\pi/4$, and $k_Fd=7$. $\gamma=2$, 6, 10, and 14 respectively from (a) to (d).}
\label{dispersion_ar}
\end{figure}
In Fig. \ref{stability}(a)-(c), we show how the velocity of collective modes, $u_j$, changes as a function of interaction strength for different
numbers of tubes. Defining $\gamma_c$ to be the critical interaction strength for the softening of the lowest excitation energy, in Fig.
\ref{stability}(d) we show that this critical value is proportional to
$k_Fd$ upto a universal function of the number of tubes, $N$.
The physical interpretation of such result is straightforward: the ratio
between $\gamma$ and $k_Fd$ is equivalent to the ratio of
interaction energy to the Fermi energy, which is the only important
parameter to determine the quantum phase diagram in our present system since the intra-tube interaction is zero here.

\subsection{Results for $\phi\neq 0$}
\label{s2.2}

\begin{figure}
\includegraphics[width=8cm]{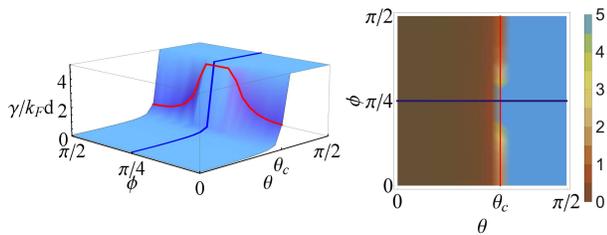}
\caption{(Color online)The critical interaction strength for the system stability as a function of dipole moment polar angle, $\theta$ and $\phi$. The number of tubes is $N=8$. We show the results both in a 3D plot and the 2D shading plot. The red and blue lines respectively the line with $\theta=\theta_c$ and $\phi=\pi/4$.}
\label{STB}
\end{figure}

When $\phi\neq0$, the dispersion relation will change in another 
fashion. In Fig. \ref{dispersion_ar}, we show the energy dispersion at $\phi=\pi/4$. At this angle, both inter- and intra-tube interaction matrix elements are zero in the long wave-length limit, i.e. 
$\tilde{V}_l(0)=0$ for $l=0,1,2\cdots$. The systems becomes different from noninteracting systems only through the interaction at finite momentum regime ($k\neq 0$). We can see that when the interaction $\gamma$ 
changes from from 2 to 14, the strong interaction separates the dispersion curves in a totally different way from $\phi=0$ case.
Different branches of excitations intersect with each other at none zero momentum, and the intersection point is almost the same even when the dipole strength increases. When the interaction is stronger than a critical value, the system becomes unstable through a softening at finite $k$ as expected. 
In Fig. \ref{STB}, we show the calculated critical interaction strength  for $N=8$, as a function of $\theta$ and $\phi$. 
We can see that the system becomes more stable when $\theta$ is larger than the magic angle, $\theta_c$, while the variation of angle $\phi$ does not affect the stability of the system very much, except for some asymmetric critical $\gamma$ between $\phi>\pi/4$ and $\phi<\pi/4$.

\section{Correlation functions and scaling exponents}
\label{s3}

Now we study the correlation functions within the Luttinger liquid theory.
The correlation functions of the order 
parameters $\hat{O}$ should decay in a power-law in the large distance limit
[\oncite{Solyom1979,Giamarchi2003,Nagaosa1999}]:
\be
\langle\hat{O}^{\da}(x)\hat{O}(0)\rangle &\sim& \displaystyle{\frac{1}{|x|^{2-\alpha}}}
\label{power}
\ee
where $\alpha$ is the associate scaling exponent. Here we have assumed that the order parameter, $\hat{O}$, here is composed by two-fermion operators only and therefore an order parameter is of quasi-long-ranged order only when $\alpha>0$. The order parameters of the largest Luttinger component should be understood as the dominant quantum phase for a given parameter. The detail calculation of the Luttinger exponent is shown in Appendix. \ref{B}. 

\begin{figure}
\includegraphics[width=8cm]{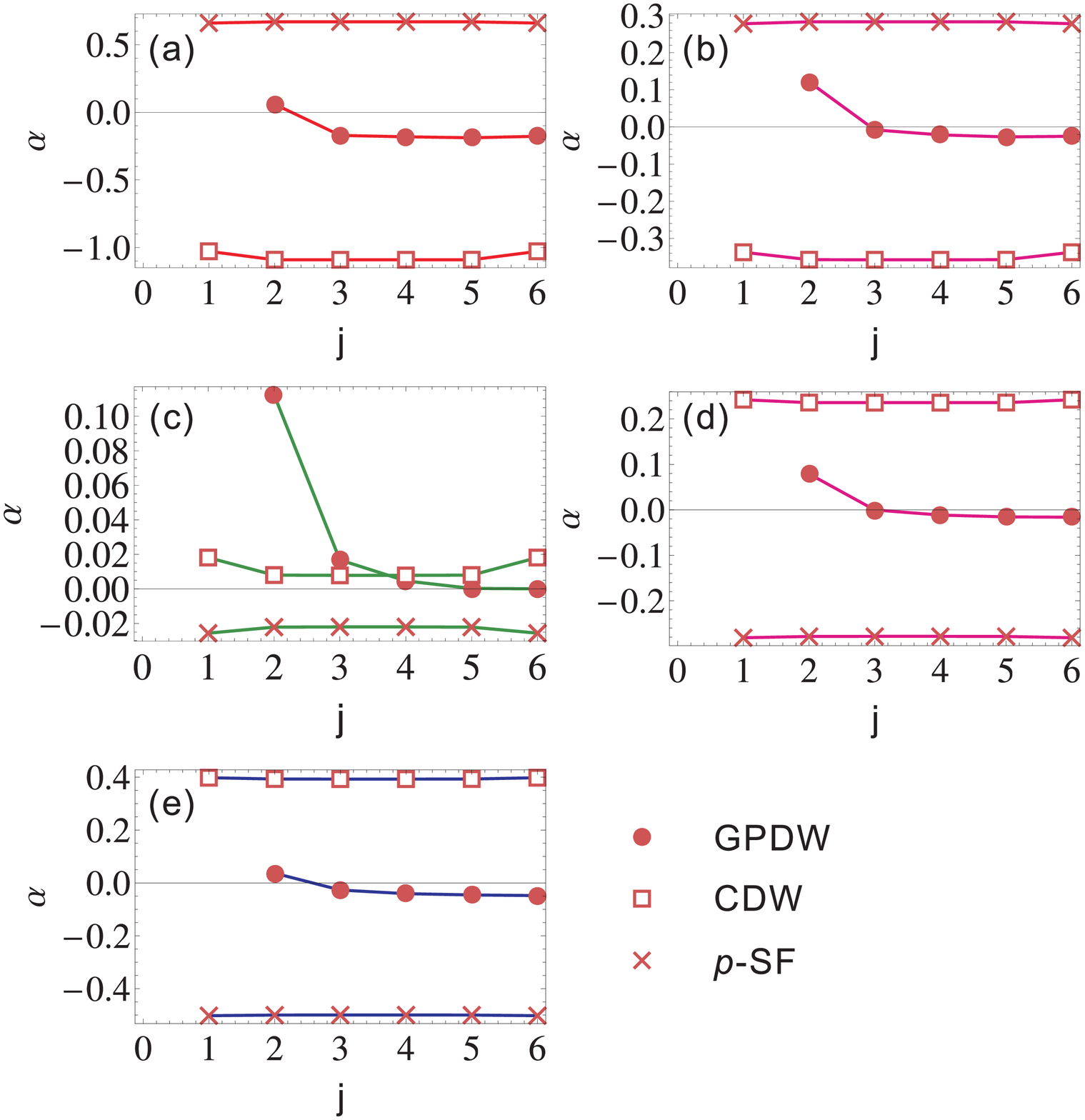}
\caption{Luttinger exponents for different order parameters at $\phi=0$. for the GPDW phase, we calculate the correlation function between the first tube and the other tubes, $j$, labeled by the horizontal index. For the other two phases, we show result of correlation function of a single tube. Here we set $\gamma=2$, $d_R=5$, $k_Fd=7$, and $N=6$. $\theta=\theta_c-2\pi/36$, $\theta_c-\pi/36$, $\theta_c$, $\theta_c+\pi/36$, and $\theta_c+2\pi/36$ respectively from (a) to (e).}
\label{correlation1}
\end{figure}
\begin{figure}
\includegraphics[width=8cm]{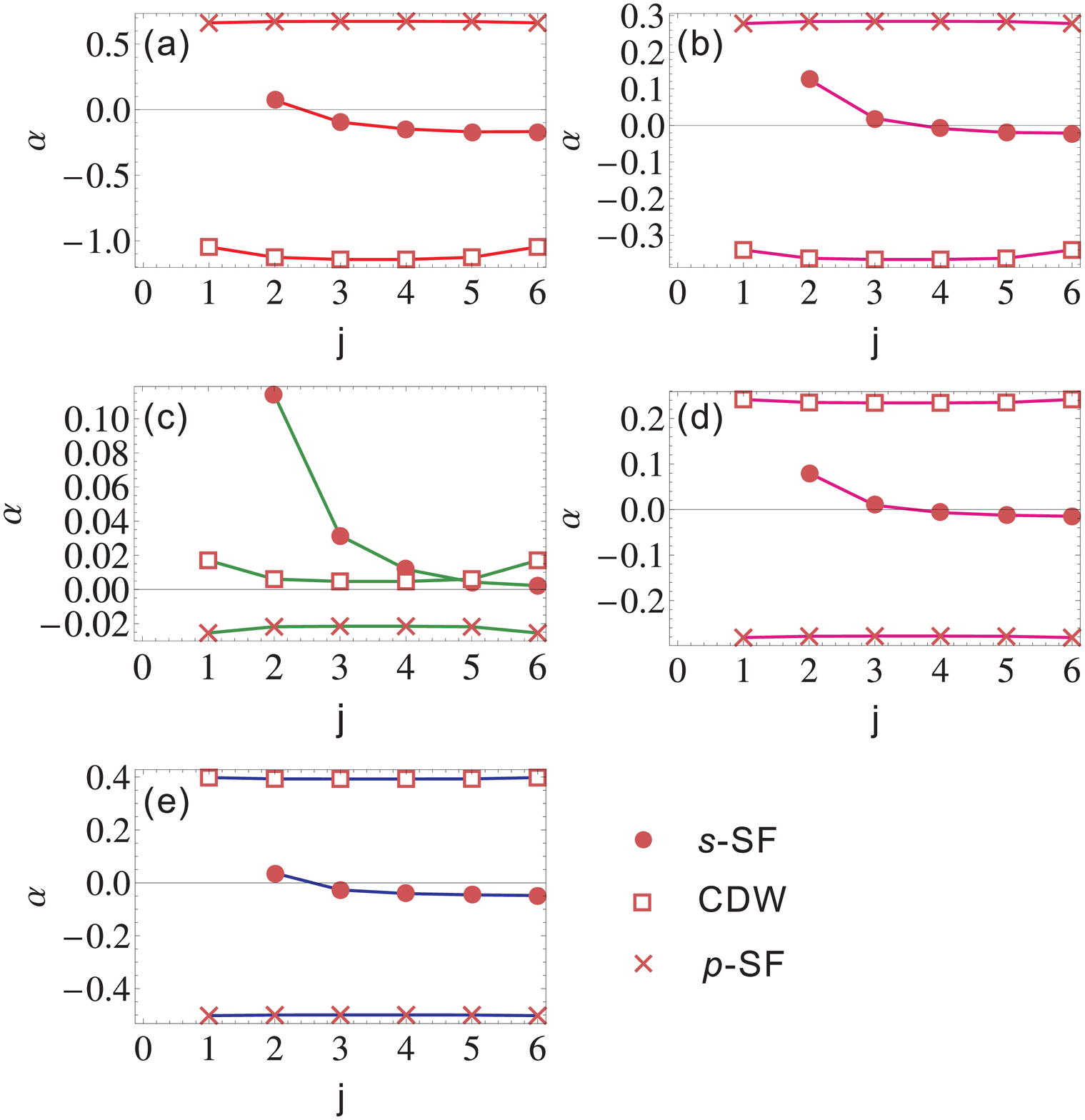}
\caption{
Luttinger exponents for different order parameters at $\phi=\pi/2$. for the $s$-SF phase, we calculate the correlation function between the first tube and the other tubes, $j$, labeled by the horizontal index. For the other two tubes, we  show result of correlation function of a single tube. Other parameters are the same as Fig. \ref{correlation1}. here $\theta=\theta_c-2\pi/36$, $\theta_c-\pi/36$, $\theta_c$, $\theta_c+\pi/36$, and $\theta_c+2\pi/36$ respectively from (a) to (e).}
\label{correlation2}
\end{figure}
Based on our earlier results in the double tube system of Ref. [\oncite{Chang2008}], here we can just consider the following four kinds of order parameters:
charge density wave (CDW), $p$-wave superfluid ($p$-SF), inter-tube gauge phase density wave (GPDW), and inter-tube $s$-wave superfluid ($s$-SF). Note that we redefine the names of some order parameters, because the usual Luttinger liquid theory has only two (pseudo-spin) components, while the case we studied here has more components. The (pseudo-)spin based terminologies (say, triplet/singlet pairing or spin density wave etc.) are therefore not suitable here at all. However, we can still define the order parameters in any pair of tubes as following:
\be
O_{\rm CDW}^{j,j'}(x) &\equiv&\psi_{j,+}^{\dagger}
(x)\psi_{j,-}(x)-\psi_{j',+}^{\dagger}
(x)\psi_{j',-}(x)
\label{orderP0}
\\
O_{\rm GPDW}^{j,j'}(x)&\equiv& \psi_{j,+}^{\dagger}(x)\psi_{j',-}(x)
+\psi_{j',+}^{\dagger}(x)\psi_{j,-}(x)
\\
O_{p-{\rm SF}}^{j,j}(x)&\equiv& \psi^\dagger_{j,-}(x)\psi^\dagger_{j,+}(x)
\\
O^{\dagger}_{s-{\rm SF}}(x)&\equiv& \psi^{\dagger}_{j,+}(x)\psi^{\dagger}_{j',-}(x)-\psi^{\dagger}_{j',+}(x)\psi^{\dagger}_{j,-}(x).
\label{orderP}
\ee
We also have used the fact that the dominant quantum phases are always contributed from fermions of two opposite chiralities ($r=\pm$). 

In Fig. \ref{correlation1}, we show the calculated Luttinger scaling exponents for each of these order parameters at $\phi=0$. Since $s$-wave superfluid never becomes a candidate of quasi-long-ranged order (i.e. its exponent, $\alpha$, is always negative in the entire range we consider here), we will not show its result for simplicity. Note that there is no correlation between fermions in tube $j$ and tube $j'$ for the CDW and $p$-SF phase, and therefore their Luttinger exponent, $\alpha$, is finite only when $j=j'$. As a result, in Fig. \ref{correlation1} we show the obtained exponent for each tube ($j=j'$) if considering these two phases, while we show results between tube $j=1$ and $j'\neq 1$ when considering the GPDW phase (similarly for the $s$-wave superfluid phase, but not shown here). 
We can find that for $\theta<\theta_c$, the dominating phase is $p$-wave superfluid, due to the strong attractive interaction between fermions in the same tube. Increasing $\theta$ gradually shows that the this phase becomes suppressed, and the inter-tube gauge phase density wave (GPDW) becomes dominate when $\theta\sim\theta_c$. For $\theta$ is getting larger, the ground state becomes dominated by the charge density wave (CDW) phase eventually. In a double tube system ($N=2$, Ref. [\oncite{Chang2008}]), such GPDW phase is also called the planar (pseudo-)spin-density wave phase, breaking the gauge symmetry of the particle conservation in each tube. In the mean field level, this corresponds to an interaction induced effective tunneling correlation between neighboring tubes, similar to the ferromagnetic state in the quantum Hall bilayer systems [\oncite{quantum_Hall}], except that the gauge phase here is not uniform, but oscillating along the tube.

In Fig. \ref{correlation2}, we show results in another limit, $\phi=\pi/2$, which makes the inter-tube interaction attractive. Again, we can see that the $p$-SF and CDW phases are the dominant phases for $\theta$ is smaller and larger than $\theta_c$. However, when $\theta\sim\theta_c$, the $s$-wave superfluid becomes dominant, showing an inter-tube pairing phase. Different from the typical BCS pairing, here the pairing is between any two nearest neighboring tubes and therefore can be regarded as a special multi-component BCS pairing phase.

\section{Complete Phase diagram and discussion}
\label{s4}

\begin{figure}
\includegraphics[width=8cm]{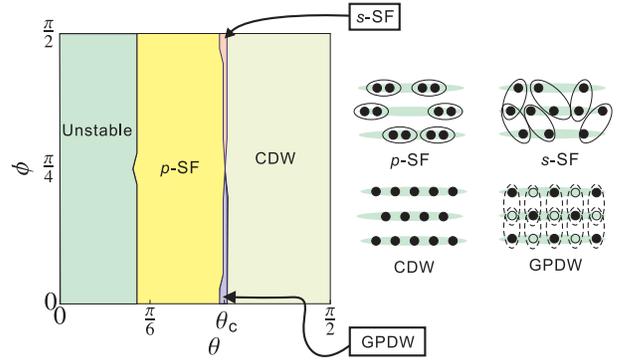}
\caption{(Coloer online) Typical quantum phase diagram for multi-tube systems for $k_Fd=7$, $d_R=5$ and $\gamma=0.6$. Right hand side are cartoons for the four dominant phases in the system. The black filled circles indicate fermions and the open circles indicate holes. The elliptic circles with solid/dashed lines indicate pairing/coherence between the two fermions.}
\label{phase_diagram}
\end{figure}

Combining above results we can summarize the ground state properties of weakly interacting dipolar fermions in a system of a planar array of 1D tubes in Fig. \ref{phase_diagram}. we also show some cartoon pictures to describe the physical meaning of each relevant order parameters in this system. We can see that by tuning the polar angle of dipole moment, we can have these four different phases (CDW, GPDW, $p$-SF, and $s$-SF) in different regime of the phase diagram. The main features to make this system different from traditional solid state systems are the appearance of inter-tube gauge-phase density wave (GPDW) and the $s$-wave superfluid ($s$-SF), where the former requires a stronger inter-tube repulsive interaction compared to the intra-tube interaction, while the later requires a stronger negative inter-tube interaction. Both of these two phases break the $U(1)$ gauge symmetry of this system (conservation of particles in each tube). 
Since the CDW phase and the intra-tube $p$-wave superfluid phase are easily understood from the nature of long-ranged repulsive and attractive interaction, here we concentrate on the physical picture and implication of the other two phases, inter-tube gauge phase density wave (GPDW) and inter-tube $s$-wave superfluid ($s$-SF) phases. 

The existence of the GPDW phase has been observed in a double-tube system in Ref. [\oncite{Chang2008}], where the pseudo-spin language is used to define it as a planar pseudo-spin density wave. As have been mentioned in Ref. [\oncite{Chang2008}], such GPDW phase breaks the $U(1)$ symmetry of the underlying Hamiltonian (i.e. particle conservation in each tube), similar to the quantum Hall ferromagnetism in a double-layer system [\oncite{quantum_Hall}]. In the quantum Hall system, the pseudo-spin ferromagnetism (also called inter-layer coherence or exciton condensate) is experimentally manifested by an interaction induced tunneling current in the limit of zero single particle tunneling. In our present system, we have such 1D inter-tube correlation between any two nearest neighboring tubes, and therefore believe that if a small (but energetically negligible) single particle tunneling is allowed, the tunneling current (quantum fluctuations) between any two nearest pair of tubes can be greatly enhanced, showing a nontrivial long-ranged correlation in the direction {\it perpendicular} to the tube direction also (see the double-tube result in Ref. [\oncite{Chang2008}]). In other words, a dipolar atom or molecule in the first tube has a finite probability to be found in the tube of the other side, even the single particle tunneling rate is much smaller than the value needed for the same result in the noninteracting case [\oncite{note1}]. Such interesting many-body effect may be also relevant to the Luttinger liquid explanation for the high-$T_c$ superconductor in the two-dimensional cuprate, proposed by P. W. Anderson [\oncite{anderson}]. We believe it is worthy to investigate further along this direction by using systems of ultracold polar molecules as a quantum simulator, which can have a very strong dipolar interaction to make the scaling exponent, $\alpha$, to be large enough to be observed experimentally.

As for the inter-tube $s$-wave superfluid, we emphasize that it is different from the usual BCS type pairing by additional multi-component physics: the Cooper pairing can occur between any two nearest neighboring tubes. For a two-component 1D system (say, spin-half electron gas), the attractive interaction between fermions also leads to the $s$-wave pairing and hence a perfect tunneling through a potential barrier at zero temperature, as first investigated by Kane and Fisher [\oncite{Kane}] by using renormalization group method. At a finite temperature, they also predicted a power-law decrease of the tunneling rate. Therefore, it is then reasonable to expect similar behavior should be also observed for the pairing between multi-component fermions. Moreover, since there are at least two choices for a fermionic dipole in the middle tube to pair another dipoles in its two neighboring tubes, such superflow property should be more significant than the result of two-component pairing case. A more extensive study of the transport property of such multi-tube systems can be very interesting and significant both in the theoretical and experimental sides, but has beyond the scope of this present paper.

\section{Summary}
\label{s5}

In summary, we study the complete quantum phase diagram of a fermionic dipolar atoms/molecules in a planar array of one-dimensional tubes.
Using the Luttinger liquid theory and the generalized Bogoliubov transformation, we are able to derive the elementary excitations and the Luttinger components for various correlation functions. 
From the mode softening of the lowest energy excitations, we find that such a fermionic dipolar system will become unstable when the interaction is stronger and when the number of tubes increases. From the calculation of Luttinger exponents, we identify the parameter regimes for various kinds of order parameters to be dominant, including charge density wave, $p$-wave superfluid, gauge phase density wave, and $s$-wave superfluid, where the last two exist only when the intra-tube interaction is smaller than the inter-tube interaction. These two interesting many-body phases cannot be realized in the semi-conductor based quantum wire systems, and are worthy to be investigated further in the future theoretical and experimental studies. 

The authors appreciate valuable discussion with H.-H. Lin, C.-M. Chang, M. Cazalilla, and E. Demler. This work is supported by NSC (Taiwan).

\appendix

\section{Luttinger scaling exponents of correlation functions}
\label{B}

In this section, we will briefly introduce the method to calculate the scaling exponents of various order parameters within the bosonization method.
We first express the fermion operator by the following bosonized expression [\oncite{Voit,Giamarchi2003}]:
\be
{\psi}_{\lambda}(x) &\approx & \lim_{\alpha\rightarrow0}\frac{e^{irk_Fx}}{\sqrt{2\pi\alpha}}\exp\left[-i(r\Phi_{\lambda}(x)-\Theta_{\lambda}(x))\right],
\label{ffield}
\ee
where $\alpha$ is a short-ranged cut-off and will be taken to be zero in the final expression of any physical quantity. We also have neglected the fermionic creation operators, which is not relevant to the calculation of correlation function here. The two new bosonic fields are defined to be
\be
\Phi_{\lambda}(x) &\equiv &
-\frac{i\pi}{L}\sum_{q\neq0}\frac{
e^{-\alpha |q|/2}e^{-iqx}}{q}\left[\rho_{j,+}(q)+\rho_{j,-}(q)\right]
\label{phi}
\nonumber\\
\\
\Theta_{\lambda}(x) &\equiv &
\frac{i\pi}{L}\sum_{q\neq0}\frac{e^{-\alpha |q|/2}e^{-iqx}}{q}\left[\rho_{j,+}(q)-\rho_{j,-}(q)\right],
\label{theta}
\nonumber\\
\ee
which are the bosonized density fluctuation and phase fluctuation operators respectively. In this section, we will use $j$ to label the different tubes and therefore $\lambda=(j,\pm)$ is to label the fluctuations of the $j$th tube with chirality $r=\pm$ (i.e. the right/left movers).
To transform the bare density fluctuation to the eigenstate bases, $\tilde{\rho}_{n}(q)$, we use the following inverse transformation for Eq. (\ref{transform1}):
\be
\rho_{j,+}(q) &=& \sum^{2N}_{n=1}T_{j,n}\tilde{\rho}_{n}(q)
\label{rtrho}
\nonumber\\
\\
\rho_{j,-}(q) &=& \sum^{2N}_{n=1}T_{j+N,n}\tilde{\rho}_{n}(q),
\label{ltrho}
\nonumber\\
\ee
where $n$ is the eigenvector index, and $n\le N$ for right moving mode ($r=+$) and $n\ge N+1$ for left moving mode ($r=-$). For symmetric model, i.e. the Fermi velocities of the left movers and the right movers are the same, the matrix, ${\bf T}$, will be block symmetric, i.e. $T_{j,n}=T_{j+N,n+N}$ and $T_{j+N,n}=T_{j,n+N}$ for $n\le N$. 
We can insert Eq. (\ref{rtrho}) and Eq. (\ref{ltrho}) into Eqs. (\ref{phi}) and (\ref{theta}) and simplify the result using the block symmetry. We obtain
\be
\Phi_{\lambda}(x) &=& \sum_{n=1}^N(T_{j,n}+T_{j+N,n})
\nonumber\\
&&\times\left\{-\frac{i\pi}{L}\sum_{q\neq0}\frac{e^{-\alpha |q|/2}e^{-iqx}}{q}[\tilde{\rho}_n(q)+\tilde{\rho}_{n+N}(q)]\right\}
\nonumber\\
&=& \sum_{n=1}^N\left(T_{j,n}+T_{j+N,n}\right)\tilde{\Phi}_{n}(x)
\equiv\sum_{n=1}^N c_n^j\tilde{\Phi}_{n}(x)
\label{nphi}
\ee
\be
\Theta_{\lambda}(x) &=&\sum_{n=1}^N(T_{j,n}-T_{j+N,n})
\nonumber\\&&\left\{\frac{i\pi}{L}\sum_{q\neq0}\frac{e^{-\alpha |q|/2}e^{-iqx}}{q}[\tilde{\rho}_n(q)-\tilde{\rho}_{n+N}(q)]\right\}
\nonumber\\&=& \sum_{n=1}^N\left(T_{j,n}-T_{j+N,n}\right)\tilde{\Theta}_{n}(x)\equiv\sum_{n=1}^N d_n^j\tilde{\Theta}_{n}(x),
\nonumber\\
\label{ntheta}
\ee
where $c_n^j\equiv T_{j,n}+T_{j+N,n}$ and $d_n^j\equiv T_{j,n}-T_{j+N,n}$.

The next step is to use the bosonized fermion field operator in Eq. (\ref{ffield}) to calculate the desired correlation function, Eqs. (\ref{orderP0})-(\ref{orderP}). We can combine all the phase fields into one exponential by using the following identities: $e^A e^B=e^{A+B}e^{[A,B]/2}$, and $\langle e^A\rangle=e^{\frac{1}{2}\langle A^2 \rangle}$ for bosonic operators. After some straightforward calculation, we find that all the correlation functions we consider in this paper can be simplified into the following forms [\oncite{Giamarchi2003}]:
\be
&&\langle O_{GDPW}^{j,j'\da}(x)O_{GDPW}^{j,j'}(0)\rangle
\nonumber\\
&\propto&\exp\left[-\frac{1}{2}\Big[\sum_n(c_n^{j'}+c_n^j)^2+(d_n^{j'}-d_n^j)^2\Big]F_1(x)\right]
\\
&&\langle O_{{\it s}-SF}^{\da}(x)O_{{\it s}-SF}(0)\rangle
\nonumber\\
&\propto&\exp\Big[-\frac{1}{2} \Big[\sum_n(c^{j'}_n-c^j_n)^2+(d^{j'}_n+d^j_n)^2\Big]F_1(x)\Big]
\\
&&\langle O_{CDW}^{j,j\da}(x)O_{CDW}^{j,j'}(0)\rangle
\propto\exp\Big[-2\left[\sum_n(c_n^j)^2\right]F_1(x)\Big]
\nonumber\\
\\
&&\langle O_{{\it p}-SF}^{j,j\da}(x)O_{{\it p}-SF}^{j,j}(0)\rangle
\propto\exp\Big[-2\Big[\sum_n(d_n^j)^2\Big]F_1(x)\Big],
\nonumber\\
\ee
where
\be
\begin{aligned}
 &F_1(x)\equiv\langle(\Phi_n(x)-\Phi_n(0))^2\rangle=\langle(\Theta_n(x)
-\Theta_n(0))^2\rangle
\end{aligned}
\label{f1}
\ee
as defined in the Ref. [\oncite{Giamarchi2003}]. Note that, we have used the fact that the CDW and $p$-SF phases are defined for fermions in the same tube ($j=j'$). The universal function, $F_1(x)$, has been calculated in details in Ref. [\oncite{Giamarchi2003}], and here we just show the final result:
\be
F_1(x)&=&\frac{1}{2}\ln \left[\frac{x^2+\alpha^2}{\alpha^2}\right].
\ee
As a result, the correlation of different order will decay as a power law of $x$:
\be
\langle O_{GDPW}^{j,j'\da}(x)O_{GDPW}^{j,j'}(0)\rangle &\propto& x^{-2+\alpha_{GDPW}}
\nonumber\\
\langle O_{CDW}^{j,j'\da}(x)O_{CDW}^{j,j'}(0)\rangle &\propto& x^{-2+\alpha_{CDW}}
\nonumber\\
\langle O_{{\it p}-SF}^{j,j\da}(x)O_{{\it p}-SF}^{j,j}(0)\rangle &\propto& x^{-2+\alpha_{{\it p}-SF}}
\nonumber\\
\langle O_{{\it s}-SF}^{j,j\da}(x)O_{{\it s}-SF}^{j,j}(0)\rangle &\propto& x^{-2+\alpha_{{\it s}-SF}}
\ee
where
\be
\alpha_{GDPW}&=&2-\frac{1}{2}\Big[\sum_n(c_n^{j'}+c_n^j)^2+(d_n^{j'}-d_n^j)^2\Big]
\nonumber\\
\alpha_{CDW}&=&2-2[\sum_s(c_s^j)^2]
\nonumber\\
\alpha_{{\it p}-SF}&=&2-2\Big[\sum_s(d_s^j)^2\Big]
\nonumber\\
\alpha_{{\it s}-SF}&=&2-\frac{1}{2} \Big[\sum_t(c^{j'}_t-c^{j}_t)^2+(d^{j'}_t+d^{j}_t)^2\Big].
\label{alpha3}
\ee
Note that for CDW and $p$-SF phases, we just need to calculate the correlation between order parameters in the same tubes, because there is no correlation between tube $j$ and tube $j'$ in the definition of order parameters. 
Therefore, after we use the generalized Bogoliubov transformation to diagonalized the bosonic Hamiltonian of the Luttinger model, Eq. (\ref{Hamiltonian}), the obtained transform matrix (Eqs. (\ref{rtrho}) and (\ref{ltrho})) can be used to calculate the Luttinger exponents directly as shown above.



\begin{thebibliography}{99}
\bibitem{note1}
More precisely, in the Luttinger liquid results derived here, this nontrivial inter-tube correlation is dominant by fermions of different chiralities with a momentum difference, $2k_F$. This is very similar to the quantum Hall ferromagnetism case with an in-plane magnetic field [\oncite{quantum_Hall,wang_QH}], which is to provide an additional gauge phase through the quantized magnetic flux for an electron tunneling from one layer to the other.
\end{thebibliography}
\end{document}